\documentstyle[aps]{revtex}
\begin{document}
\draft
\title{MULTIBAND  MODEL  OF  HIGH Tc  SUPERCONDUCTORS}
\author{R. COMBESCOT} 
\address{Laboratoire de Physique Statistique,
 Ecole Normale Sup\'erieure*,
24 rue Lhomond, 75231 Paris Cedex 05, France}
\date{Received \today}
\maketitle

\begin{abstract}
We propose an extension to other high $T_{c } $  compounds of a  
model introduced earlier for YBCO. In the ''self-doped'' compounds  
we assume that the doping part ( namely the BiO, HgO, TlO planes  
in BSCCO, HBCCO, TBCCO respectively ) is metallic, which leads  
to a multiband model. This assumption is supported by band structure  
calculations. Taking a repulsive pairing interaction between  
these doping bands and the $CuO_{2} $ bands leads to opposite  
signs for the order parameter on these bands and to nodes whenever  
the Fermi surfaces of these bands cross. We show that in BSCCO  
the low temperature dependence of the penetration depth is reasonably  
accounted for. In this case the nodes are not located near the  
$45^{o}$ direction, which makes the experimental determination of  
the node locations an important test for our model. The situation  
in HBCCO and TBCCO is rather analogous to BSCCO. We consider  
the indications given by NMR and find that they rather favor  
a metallic character for the doping bands. Finally we discuss  
the cases of NCCO and LSCO which are not ''self-doped'' and where  
our model does not give nodes.
\end{abstract}

\section{INTRODUCTION} 
The recent years have seen a marked progress in the debate on  
the mechanism of high $T_{c } $  superconductivity. Indeed the  
existence of nodes and change of sign of the order parameter  
at the Fermi surface has found a very important experimental  
support in some compounds like $YBa_{2}Cu_{3}O_{7} $ (YBCO)  
. This includes on one hand various experiments showing the  
existence of low energy excited states, such as tunneling,  
Raman scattering and penetration depth \cite{leggett}.  
Moreover more spectacular experiments designed to probe the  
change of sign have given positive answers. These are the corner  
SQUID experiments \cite{1} and the observation of the half magnetic  
flux quantum in 3 grain-boundary Josephson junctions \cite{2}.  
This change of sign is a clear indication that there is some  
important repulsive contribution in the pairing interaction.  
Since the spin fluctuation mechanism \cite{6} is based uniquely  
on a repulsive interaction, it appears as a leading candidate  
for the pairing mechanism in these high $T_{c } $ superconductors.  
In its simplest version it leads to an order parameter with  
a $d_{x^{2}-y^{2}} $ symmetry, commonly called d-wave order parameter.  \par 
  \bigskip 
Unfortunately this simple spin fluctuation interpretation meets  
some problems in YBCO. An important one is the weak sensitivity  
of the critical temperature of $YBa_{2}Cu_{3}O_{7} $  to defects.  
Any kind of impurities produces in d-wave superconductors an  
effect analogous to pair-breaking by magnetic impurities in  
standard s-wave superconductors, leading to a rapid decrease  
of the critical temperature with increasing scattering rate,  
as measured for example by the residual resistivity. This is  
in contrast with some experiments showing a decrease of $T_{c } $  
an order of magnitude slower than expected \cite{sun}.   \par 
  \bigskip 
Another problem is the marked anisotropy in the a-b plane shown  
by a number of recent experiments in YBCO. The existence of  
a sizeable Josephson current
 \cite{sunjo} in a c-axis $Pb-YBa_{2}Cu_{3}O_{7} $  
tunnelling junction shows that pure d-wave symmetry is not satisfied.  
Because of the large size of this Josephson current as compared  
to what would be expected from a superconductor with s-wave  
symmetry, it is difficult to believe that the departure from  
pure d-wave is only due to the small orthorhombic 1,6\% distorsion  
of the $CuO_{2} $  planes. Various other properties have also  
shown a clear in-plane anisotropy. The resistivity has an anisotropy  
of about a factor 2. Quite recently a sizeable anisotropy of  
the thermal conductivity has been observed \cite{gagnon} which  
reaches 30\% in the superconducting state.   \par 
  \bigskip 
The anisotropy of the penetration depth is perhaps the most  
striking one since it is directly related to the nature of the  
superconducting state. Indeed \cite{basov} at zero temperature,  
the ratio between the superfluid densities 
$\lambda _{b}^{-2}(0)/\lambda _{a}^{-2}(0) $   
is 2.3 , which is in good agreement with the anisotropy of the  
normal state resistivity and of the square of the plasma frequency  
\cite{basov}. Moreover we have also to take into account that  
in $YBa_{2}Cu_{4}O_{8} $ , which has twice as many chains as  
$YBa_{2}Cu_{3}O_{7} $  and an orthorhombic distorsion of 0.8\%  
, the superfluid density anisotropy \cite{basov} is as high  
as 6.2 . This trend points toward the natural conclusion that  
the chains are conducting and that the superfluid density anisotropy  
is due to their direct contribution. Now we attribute the linear  
term at low temperature to nodes in the gap. 
Since in $\lambda _{b}^{-2}(T) $   
this linear term is typically 3 times the linear 
term in $\lambda _{a}^{-2}(T), $   
the nodes contribution is also strongly anisotropic. This leads  
to the suspicion that chains have also something to do with  
the nodes. This is not so easy to include in the spin fluctuation  
picture because it deals only with the $CuO_{2} $ planes. This  
gives rather a strong hint that we have to include chains explicitely  
in our modeling of the superconducting state in order to come  
up naturally with such an anisotropy. Another strong indication  
that chains play an important role in the superconducting properties  
comes from very recent penetration depth measurements \cite{srikanth}  
which are partially in contradiction 
with earlier results \cite{hardy93},  
and show a kink in the temperature dependence. This seems to  
indicate that a two component model is appropriate for the description  
of superconductivity and leads again to a two band model. Naturally  
the present overall experimental picture is still somewhat uncertain  
and future experiments will help to clarify the situation.   \par 
  \bigskip 
These difficulties, as well as others, with the spin fluctuation  
mechanism have led us to propose recently a model \cite{10}  
which allows to account for all the above experimental facts.  
In this model pairing in the planes is due to an attractive  
interaction, while Coulomb repulsion induces in the chains an  
order parameter with opposite sign. It is well known that these  
plane and chain bands hybridize because electrons are physically  
allowed to jump from planes to chains. Due to the anticrossing  
produced by this hybridization, one obtains automatically an  
order parameter which changes sign on a single sheet of the  
Fermi surface and has nodes in the gap in the region of the  
Brillouin zone where the anticrossing takes place. In this sense  
our order parameter is d-wave like. However in contrast with  
the $d_{x^{2}-y^{2}} $ order parameter, there is no symmetry breaking  
and we have a total of 8 nodes ( 4 on each sheet of the Fermi  
surface ) instead of 4. Since the chains play an essential role  
in our model, it is meaningless to consider an approximate tetragonal  
symmetry. Rather we have to consider only the orthorhombic symmetry,  
under which our order parameter is completely invariant (no  
symmetry breaking), which is also the case for the $d_{x^{2}-y^{2}} $   
order parameter . We can also say that on each sheet  
of the Fermi surface our order parameter looks qualitatively  
like a d+s order parameter.  \par 
  \bigskip 
Since the order parameter of our model has nodes, it accounts  
naturally for all the experiments providing evidence for its  
change of sign and for the existence of low energy excitations.  
At the same time it provides also a simple explanation  
for the important anisotropy in the ab plane mentionned above,  
because of the important role of the chains. We have found \cite{11}  
that the existence of two weakly coupled bands (plane and chain)  
in our model, as well as weak plane-chain scattering, leads  
to the possibility of a weak sensitivity of the critical temperature  
to impurities, in agreement with experiment. We find \cite{lambda}that  
our model accounts quite well for the anisotropy of the penetration  
depth and for the absolute values. We are also able to reproduce  
fairly well the whole temperature dependence for both the  a  
 and the  b  directions found in Ref.\cite{hardy93}, including  
the linear dependence at low temperature.  \par 
  \bigskip 
Although our model is quite satisfactory for YBCO, in part because  
there are strong experimental indications that chains play a  
significant role, it appears for the same reason quite specific  
of YBCO and does not seem to be applicable to other compounds.  
However the natural tendency is to find a universal explanation  
for high $T_{c } $ superconductivity and hence a model applicable  
to all cuprate superconductors, since they all share fairly  
high $T_{c } $ and $CuO_{2} $ planes. 
On this basis our model sounds  
quite unsatisfactory. Let us merely remark that fortunately  
this universality hypothesis can be checked and hopefully will  
be checked experimentally. We note also that the available data  
do not seem to confirm completely this hypothesis, 
since $Nd_{2-x }Ce_{x }CuO_{4} $   
(NCCO) on one hand and YBCO (for the strong anisotropy discussed  
above) on the other do not fit in the picture. The case of NCCO  
will be discussed below. It would clearly not be satisfactory  
to find models specific to each of the different high $T_{c } $  
compounds, but it is possible that, while sharing common features,  
they also differ in some respect. A related unappealing feature  
of our model is its complexity, since it is basically a two  
band model, not as nice as a one band model. Ideally one would  
certainly rather like to have a simple rather than a complicated  
model. However we already know that these compounds are chemically  
complicated. The same might unfortunately be true for their  
physics.  \par 
  \bigskip 
Whatever the feeling about the above issues, there remains a  
more practical question. Other high $T_{c } $ compounds have been  
investigated experimentally and have shown features similar  
to the ones found in YBCO ( although the experimental picture  
is not as complete as in YBCO). Is it possible to extend the  
ideas of our model and find satisfactory explanations for them  
? And more generally, following these ideas, what do we expect  
to find in other compounds ? It is the purpose of this paper  
to consider this question. As we mentionned our model for YBCO  
is basically a multiband model. Therefore we will in particular  
consider whether a multiband model is applicable to a given  
compound and if this can lead to reasonable explanations for  
experimental facts. We note that a multiband model applies trivially  
in all the compounds having more than one $CuO_{2} $ plane in  
the elementary cell. The usual trend is to consider that this  
leads to almost degenerate bands and to treat them as a single  
band. It is not completely obvious that this is correct, but  
anyway this will not be our main emphasis. We will rather be  
interested in looking if there are bands, analogous to the chain  
band of YBCO, that is not linked to the $CuO_{2} $  planes, which  
play physically an important role. Naturally we do not intend
to review the whole situation but only to make our best to  
discuss what we believe to be relevant points. Our conclusion  
is that we can extend our model in a natural way and we do not  
find major incompatibility with experiments. Naturally this  
is a somewhat imprecise statement and we will be more specific  
below. We consider successively the cases of BSCCO, HBCCO, TBCCO,  
and finally LSCO and NCCO.  \par 

\section{BSCCO} 
We turn first to the extension of our model 
to $Bi_{2}Sr_{2}CaCu_{2}O_{8+\delta } $   
(BSCCO), which is probaby the second most investigated high  
$T_{c } $ superconductor after YBCO. This is not 
a very good compound  
for fundamental investigations since it 
is generally not stoechiometric  
but rather has often an oxygen ''doping'' $\delta  $ 
giving the maximum  
$T_{c } $ . This comes in addition to the incommensurate distorsion  
arising in the BiO planes. This is in contrast with YBCO where  
the nearly stoechiometric compound has been mostly investigated.  
We first note the high value of  $2\Delta /T_{c } $ 
(which is of order  
7 or even higher, as obtained from tunnelling experiment \cite{renner}  
as well as from Raman scattering \cite{staufer}) gives an indication  
that a simple d-wave model does not properly describe BSCCO.  
This value is incompatible with a weak coupling d-wave model,  
and it is doubtful that strong coupling spin fluctuations calculations  
\cite{pines} can lead to such a high value, because the spectrum  
for spin fluctuations goes to high energy and the coupling constant  
is of order unity. A related difficulty is the value of the  
slope of the penetration depth $\lambda (T) $ at low 
temperature \cite{jacobs,lee}. In the $\mu  $ model of Xu et al. 
\cite{xu}, which is an extension  
of the simple d-wave model allowing for a variable slope for  
the opening of the gap, $\Delta (\theta ) $ = $\Delta _{0} $ 
$\mu  $ $\theta  $  for  0 $< $ $\theta  $  
$< $ $\mu ^{-1} $ and $\Delta (\theta ) $ = $\Delta _{0} $
for  $\mu ^{-1} $ $< $ $\theta  $ $< $  
$\pi /4 $  $. $  The low temperature $\lambda (T) $ 
is given by \cite{xu}:  
[ $\lambda (T) $ - $\lambda (0) $ ] / $\lambda (0) $ 
= ( 2 ln2 / $\mu  $ $\Delta _{0} $ ) T . Taking  
the experimental value $\lambda (T) $ - $\lambda (0) $ 
$\approx  $ 10 \.A / K  and $T_{c } $  
$\approx  $ 90 K , this leads to  $\mu  $ $\Delta _{0} $ 
/ $T_{c } $ = 1.5  $10^{-3} $   
 $\lambda (0) $ , where $\lambda (0) $ is measured in \.A . 
There is some uncertainty   
in $\lambda (0). $ If we take the assumed values of 
2100 \.A \cite{lee}  
and  2600 \.A \cite{jacobs}, we have either 
$\mu  $ $\Delta _{0} $  / $T_{c } $   
$\approx  $ 3.2  or $\mu  $ $\Delta _{0} $  / $T_{c } $ 
$\approx  $ 4 . Taking 
the ( pretty unphysical  
) lower limit $\mu  $ = 4 $/\pi  $ leads to  $2\Delta _{0} $ 
/ $T_{c } $ = 5  or  
 6.2 , somewhat below experiment. The standard d-wave corresponds  
essentially to $\mu  $ = 2 and gives $2\Delta _{0} $ / $T_{c } $ = 3.2  or  
 4 , barely compatible with weak coupling and in clear disagreement  
with experiment. It might naturally be that further experiment  
give a higher value for $\lambda (0) $ 
and a lower one for $2\Delta /T_{c } $   
 which would solve this difficulty.  \par 
  \bigskip 
The natural way to extend our two-band model to BSCCO is to  
assume that the BiO planes lead to a metallic band, which would  
play the role of the chain-band in YBCO. The existence of this  
band at the Fermi level is controversial. There is some experimental  
evidence from photoemission that BiO planes are insulating because  
the BiO band is not seen \cite{dessau}, but these experiments  
are somewhat contradictory \cite{wells}. Similarly STM seems  
to indicate that the BiO plane at the surface is nonmetallic  
 \cite{renner}. However both kind of experiments test only  
the surface of the sample on a few tens of \.A, that is basically  
one cell depth. Just because the surface necessarily implies  
some kind of lattice distorsion and/or consequently some modification  
in the electronic spectrum, the BiO planes may well be insulating  
at the surface while they are conducting in the bulk. It may  
also very well be that there is some oxygen depletion at the  
surface. An example of this kind of problem is found in YBCO  
\cite{picksc}. In this compound the bands corresponding to $CuO_{2} $  
planes are seen in photoemission, while the CuO chain-derived  
band is not seen. This could lead again to the conclusion that  
this chain band is nonmetallic. However this chain band is seen  
by ACAR (which does not see the plane bands) which shows that  
this band is metallic. As indicated above this is confirmed  
completely independently by the in-plane anisotropy of the conductivity  
and mostly by the strong anisotropy of the superfluid density.  
Other problems arising with the interpretation of photoemission  
experiments are discussed in \cite{leggett}.  \par 
  \bigskip 
The main reason to believe that the BiO planes are metallic  
comes from band structure calculations \cite{krak,massbi}. Indeed  
these papers find that, in addition to the two nearly degenerate  
$CuO_{2} $ plane-derived bands, there is at the Fermi level a BiO-derived  
band. The plane bands are rather similar to the ones found in  
YBCO, and the corresponding Fermi surface looks roughly as a  
circle centered at  $(\pm  $ $\pi  $ / a , $\pm  $ 
$\pi  $ / a ) if one uses the simplified  
Brillouin zone corresponding to tetragonal $CuO_{2} $ planes.  
The BiO band gives a smaller Fermi surface centered at  ( $\pm  $  
$\pi  $ / a , 0 ) and ( 0 , $\pm  $ $\pi  $ / a ). Naturally one may wonder  
about the validity of band structure calculations. Clearly there  
are limitations in the precision of these complicated calculations.  
They are also sensitive to the precise location of the atoms.  
And LDA, which is commonly used to take into account correlations,  
is not an exact method for dealing with highly correlated systems.  
As a result mass renormalization due to interactions is not  
included and the calculations provide essentially the bare excitation  
spectrum (note however that mass renormalization due to Coulomb  
interaction seems to be rather small, from all the calculations  
dealing with correlation effects). On the other hand the results  
from different authors are in fair agreement, which indicates  
at least that the results are reasonably well converged. Moreover  
in the case of YBCO, the results are in reasonable agreement  
with all experimental informations from ARPES, ACAR and dHvA  
\cite{picksc}, including a small electron pocket centered at  
$(\pm  $ $\pi  $ / a , $\pm  $ $\pi  $ / a) , 
which is apparently not of physical   
importance. It is therefore difficult to ignore these results  
and one has to consider seriously the possibility that their  
conclusion is correct.  \par 
  \bigskip 
In the case of BSCCO the BiO band is mostly above the Fermi  
level, but the bottom of the band is located around 0.5eV below  
the Fermi level. It is likely that the precision of the method  
is better than this, and the BiO band will not be pushed above  
the Fermi level by some imprecision in the calculations. Therefore  
the existence of this BiO Fermi surface seems rather secure.  
On the other hand the calculations show a second BiO band which  
comes quite close to the Fermi level from above (within 0.1eV  
or less), but does not cross it. It is not obvious that the  
precision is good enough to make sure that there is not a second  
BiO sheet of the Fermi surface. Experimentally, as mentionned  
above, ARPES does not see the BiO band, but it sees the $CuO_{2} $  
bands ( or at least one of them, but the failure to see two  
bands might be due to their splitting being very small ). The  
Fermi surface corresponding to this band is in good agreement  
with band structure calculations, which gives again confidence  
in these calculations. On the other hand ACAR \cite{chan} gives  
results in fair agreement with band structure calculations and  
sees the BiO piece of the Fermi surface, although its size is  
somewhat larger than the theoretical prediction and the separation  
between BiO and $CuO_{2} $ Fermi surfaces larger than predicted.  
Recent experiments \cite{marel} on the mid-infrared reflectivity  
in $Bi_{2}Sr_{2}CuO_{6} $ have also been interpreted in terms  
of transitions between BiO and $CuO_{2} $ type bands. As mentionned  
above the discrepancy with ARPES might be due to the fact that  
the surface structure is different from the bulk. A similar  
problem may be seen in $Sr_{2}RuO_{4}, $ a layered perovskite  
becoming superconducting around 1K. Here there is some disagreement  
between ARPES \cite{lu} and dHvA \cite{macken} experiments,  
and it has been suggested \cite{mazin}that the origin might  
be surface relaxation of the oxygen position.  \par 
  \bigskip 
Actually calling this electron pocket the BiO piece of the Fermi  
surface is a misnomer. Indeed the corresponding electronic states  
are \cite{krak,massbi} strong hybrid of states describing electrons  
moving respectively in BiO planes and in $CuO_{2} $ planes. The  
situation is quite similar to the one found in YBCO where we  
start with electrons belonging either to planes or chains, and  
which are hybridized due to the hopping of an electron from  
plane to chain, through the apical oxygen. In the same way we  
can start in BSCCO with ''non-interacting'' bands, as considered  
explicitely by Massida et al. \cite{massbi}, corresponding to  
electrons moving either in BiO planes or in $CuO_{2} $ planes.  
Then hopping through the apical oxygen produces an hybridization  
of these bands \cite{massbi}. The net result is that \cite{massbi}  
the BiO planes contribution to the total density of states at  
the Fermi level is almost half of the one of the $CuO_{2} $  planes,  
which shows clearly the importance of BiO planes.   \par 
  \bigskip 
It is possible to produce semiquantitatively the results of  
band structure calculations with a simple tight-binding model.  
Let us indeed consider that BSCCO is build by stacking up sets  
made of two $CuO_{2} $  planes and two BiO  planes. Motion  
in the $CuO_{2} $  planes and in the BiO  planes is described  
by the dispersion relations  $\epsilon _{k } $ and  
$\epsilon ^{\prime}_{k } $ respectively.  
Moreover we allow hopping toward the nearest plane. We call  
 t  the hopping term between a $CuO_{2} $  plane and the nearest  
BiO  plane, t' the hopping term between the two BiO  planes
and t'' the one between the two $CuO_{2} $  planes, assuming for  
simplicity that they are wavevector independent. The corresponding  
Hamiltonian is: 
\begin{eqnarray}
\label{eq1}
\matrix{{H}_{0}=\sum\nolimits\limits_{n}^{}
{\varepsilon }_{k}{c}_{k,1,n}^{+}{c}_{k,1,n}+\sum\nolimits\limits_{n}^{}
{\varepsilon }_{k}{c}_{k,2,n}^{+}{c}_{k,2,n}+\sum\nolimits\limits_{n}^{}
{t"}^{}({c}_{k,1,n}^{+}{c}_{k,2,n}+h.c.)\cr \cr +\sum\nolimits\limits_{n}^{}
{\varepsilon '}_{k}{d}_{k,1,n}^{+}{d}_{k,1,n}+\sum\nolimits\limits_{n}^{}
{\varepsilon '}_{k}{d}_{k,2,n}^{+}{d}_{k,2,n}+\sum\nolimits\limits_{n}^{}
{t'}^{}({d}_{k,1,n+1}^{+}{d}_{k,2,n}+h.c.)\cr \cr
 +\sum\nolimits\limits_{n}^{}
{t}^{}({c}_{k,2,n}^{+}{d}_{k,1,n}+h.c.)+\sum\nolimits\limits_{n}^{}
{t}^{}({c}_{k,1,n}^{+}{d}_{k,2,n}+h.c.)\cr}
\end{eqnarray}
where $c_{k }^{+} $ and $d_{k }^{+} $ are creation operators in  
the $CuO_{2} $  planes bands and in the BiO  planes bands respectively.  
The indices 1 and 2 number the $CuO_{2} $  planes and BiO  planes,
while the index n  numbers the stacks. Actually we can simplify  
further by taking into account that BSCCO is very anisotropic  
which implies that some hopping terms are quite small. From  
band structure calculations \cite{krak} it is clear that hopping  
between $CuO_{2} $  planes and between BiO planes are small,  
while hopping between neighboring $CuO_{2} $  plane and BiO plane 
are somewhat stronger, leading to rather isolated $CuO_{2} $   
- BiO stacks. For simplicity we assume for example hopping between  
BiO  planes negligible, that is t' = 0, and t'' $\not=  $ 0 . But  
we could take the opposite assumption t' $\not=  $ 0, and t'' = 0 and  
reach the same qualitative conclusions. The important point  
is that the sets made of two $CuO_{2} $  planes and two BiO  planes
are now independent and we have a two-dimensional problem. The  
Hamiltonian can then be separated in two independent parts $H_{\pm } $  
 corresponding to even and odd combinations. Introducing the  
operators $c_{\pm } $  and $d_{\pm } $  by  $c_{1,2} $ 
= ( $c_{+} $ $\pm  $ $c_{-} $   
$)/\surd 2 $  and $d_{1,2} $ = ( $d_{+} $ 
$\pm  $ $d_{-} $ $)/\surd 2 $ : 
\begin{eqnarray}
\label{eq2}
{H}_{\pm }=\sum\nolimits\limits_{k}^{}
({\varepsilon }_{k}\pm
{t"}^{}){c}_{k,\pm}^{+}{c}_{k,\pm}+\sum\nolimits\limits_{k}^{}
{\varepsilon'}_{k}{d}_{k,\pm}^{+}{d}_{k,\pm}+
\sum\nolimits\limits_{k}^{}
{t}^{}({c}_{k,\pm}^{+}{d}_{k,\pm}+h.c.)
\end{eqnarray}
Each of these Hamiltonian is analogous to the two-dimensional  
model \cite{10} we have considered for YBCO. For each of them  
the $CuO_{2} $  band and the BiO band hybridize. Since band  
structure calculations show that one of the BiO derived band  
is pushed above the Fermi level by this coupling, we consider  
only the other Hamiltonian, where hybridization  
leads to an anticrossing of the Fermi lines. In order to obtain  
a Fermi surface in reasonable agreement with band structure  
calculations, we take for the unhybridized $CuO_{2} $  band the  
rather standard dispersion relation : 
$\epsilon _{k } $  - t'' = - $2t_{0} $   
( $\cos(k_{x}a) $ + $\cos(k_{y}a) $ )
+ $2t_{0} $ $\cos(k_{x}a) $ $\cos(k_{y}a) $   
- $\mu  $  $    $ with $t_{0} $  = 0.33 eV, 
$\mu  $  $   = $ - 0.46 eV (the axes   
are along the CuO  bonds). On the other hand the unhybridized  
BiO band can be described by a nearest neighbour tight binding  
approximation (with BiO bonds rotated by $45^{o}$ with respect to  
the CuO  bonds), which leads to 
$\epsilon ^{\prime}_{k } $  = $4t'_{0} $ 
$\cos(k_{x}a) $ $\cos(k_{y}a) $ - $\mu ^{\prime} $ ,
with  $t'_{0} $  = 0.3 eV and $\mu ^{\prime} $  = 
- 0.6 eV  in order to reproduce band structure calculations.  
The result for the Fermi lines is shown in Fig.1 for t = 0.1  
eV . This figure is very similar to the Fermi surfaces found  
in \cite{krak,massbi}. We can then take over our model \cite{11}  
for YBCO and assume that there is a repulsive pairing interaction  
between $CuO_{2} $  plane and BiO  plane, leading to an order  
parameter with opposite sign in the $CuO_{2} $  plane and BiO  
plane. Then because of hybridization we have nodes in the  
anticrossing regions. To give an example we assume the regime  
of well separated bands, where the hopping term  t  is large  
compared to the gaps (actually it is not clear that this condition  
is satisfied because the maximum gap is quite large in BSCCO  
as we have seen above, but this should not lead to qualitative  
changes). In this case, assuming for simplicity isotropic order  
parameters $\Delta  $ in the $CuO_{2} $  plane 
and $\Delta ^{\prime} $ in the BiO plane, 
the order parameter is given \cite{11} by  
$\Delta _{k } $ = ( $\Delta  $ $\epsilon _{k }^{\prime} $   
+ $\Delta ^{\prime} $ $\epsilon _{k } $ )
/( $\epsilon _{k } $ + $\epsilon _{k }^{\prime}) $ 
when hybridization is  
taken into account. In particular the locations of the nodes  
is not so sensitive to the ratio $\Delta /\Delta ^{\prime} $ 
because the Fermi lines   
for unhybridized $CuO_{2} $  plane and BiO  plane cross essentially  
at right angle. We show in Fig.1 the locations of the nodes  
for  $\Delta /\Delta ^{\prime} $ = - 1 .  \par 
  \bigskip 
We can then try to calculate the temperature dependence of the  
penetration depth, just as we have done for YBCO \cite{lambda}.  
However the results depend sensitively on a number of ingredients  
\cite{lambda} and the situation is here much more uncertain  
than in YBCO, in particular with respect to the band structure  
and the parameters describing the BiO derived bands. Therefore  
we will only consider the low temperature behaviour where the  
number of hypotheses can be somewhat more limited. We proceed  
as in \cite{lambda} and treat the hopping term   t  as a small  
quantity so that hybridization modifies the dispersion relations  
only in the vicinity of the crossing of the unhybridized Fermi  
lines. This leads to : 
\begin{eqnarray}
\label{eq3}
{\lambda }^{-2}(0)-{\lambda }^{-2}(T)={4 \ln 2 \over \pi }
{{e}^{2}{\mu }_{0}\over {\hbar}^{2}}{t \over c}{T \over {(\Delta
\left|{\Delta'}\right|)}^{1/2}}{A}_{0}
\end{eqnarray}
where $\Delta  $ and $\Delta ^{\prime} $ are the order 
parameters in the unhybridized   
$CuO_{2} $ and BiO planes respectively (which we have assumed  
to be {\bf k}-independent for simplicity) and c = 14.6 \.A is  
the thickness of the stack made of two $CuO_{2} $  planes and  
two BiO  planes. The dimensionless constant $A_{0} $  is related  
to the quasiparticle velocities at the crossing point of the  
unhybridized Fermi lines. We assume for simplicity that these  
velocities are orthogonal (this is essentially the case in Fig.1).  
In this case $A_{0} $  is given by $A_{0} $  = $(\Delta ^{2} $ $v_{B}^{2} $  
+ $\Delta ^{\prime   2} $ $v_{C}^{2} $ $)/
(\Delta  $ $|\Delta ^{\prime}| $ $v_{B} $ 
 $v_{C}) $ where $v_{C} $  
 and $v_{B} $  are the velocity of the unhybridized $CuO_{2} $ and  
BiO bands respectively. This quantity has a minimum equal to  
2 when  $\Delta  $ $v_{B} $ = $|\Delta ^{\prime}| $ $v_{C} $ 
and we take for simplicity   
this worst case for our evaluation. Expressing numerically all  
known physical constants, this can be rewritten in the low T  
regime as : 
\begin{eqnarray}
\label{eq4}
{\lambda }^{}(T)-{\lambda }^{}(0)=2.8{\lambda }^{3}(0){T \over
{T}_{c}}{{T}_{c} \over {(\Delta \left|{\Delta '}\right|)}^{1/2}} t
\end{eqnarray}
where the all penetration lengths are now expressed in units  
of  1000 \.A  and  the hopping parameter  t  is in eV. We want  
now to compare this result with experiments \cite{jacobs,lee}  
on BSCCO which gives a slope of 10 \.A/K. We take for example  
$\Delta /\Delta ^{\prime} $ = - 1 (actually we 
would expect $\Delta ^{\prime} $ to be somewhat less  
than  $\Delta , $ which would be make our case easier), and $\lambda (0) $ =  
2000 \.A from experiment (again a rather worst case hypothesis).  
If we use $\Delta /T_{c } $ $\approx 3 $ 
as discussed above, we obtain agreement   
with experiment for t = 0.12 eV . This result is quite similar  
to what we have found for YBCO. It is small enough to make our  
hybridization approach consistent. This is quite satisfactory  
since we have basically no adjustable parameters. On the other  
hand it is not in contradiction with the strong anisotropy and  
the large resistivity well known in BSCCO, since this can be  
attributed entirely to the very small hopping between either  
the BiO planes or the $CuO_{2} $ planes or both, as we have already  
discussed. It may seem rather strange that we can account for  
the rather strong slope found experimentally by a small hybridization.  
Roughly speaking this is due to the fact that we have a total  
of 16 nodes in the Brillouin zone, whereas the standard d-wave  
model has only 4 nodes.  \par 
   \par 

Our model differs clearly from standard d-wave, not only by  
the number of nodes but also by their locations. This is in  
contrast with YBCO where the location of the crossing between  
plane and chain bands is not so far from the $45^{o}$ direction of  
the standard d-wave model. Here,  in Fig.1 , the directions  
of the nodes are respectively $18^{o}$ and $22^{o}$ (and all the other  
ones obtained by tetragonal symmetry), with the origin taken  
at $(\pi ,\pi ) $ as it is usually done. This is quite different from  
the $45^{o}$ of the d-wave model. Naturally Fig.1 is just an example,  
and it is likely that the situation in actual BSCCO is somewhat  
different because band structure ( as suggested by ACAR experiments)  
and order parameter are not the ones we have chosen. However  
we do not expect the location of the nodes to change qualitatively.  
Therefore experiments designed to ascertain the position of  
the nodes are quite important and we consider now briefly the  
evidence for this position in BSCCO (we will also discuss this  
point below for HBCO). At first photoemission experiments were  
interpreted \cite{ding} as showing nodes at $35^{o}$. Taking together  
into account experimental and theoretical uncertainties, we  
would not consider this as being in clear contradiction with  
our model. However these experiments have then been reinterpreted  
\cite{ding1} as giving the nodes at $45^{o}$ which agrees with d-wave  
and disagrees with our model. Anyway, as we discussed above,  
it is not clear at all that photoemission provides a faithfull  
picture of what is happening in the bulk. And it sees only a  
$CuO_{2} $  derived band, and no BiO band. So the fact that our  
model disagrees with photoemission is not so surprising.   \par 
  \bigskip 
On the other hand in-plane tunneling experiments give an indication  
on the location of the nodes. They have shown a clear gap anisotropy  
\cite{kane}with a minimum gap along the CuO bond direction in  
contrast with photoemission experiments. Naturally, just as  
photoemission, this technique is limited in angular resolution  
and can not by itself display nodes, but the obvious conclusion  
is that the nodes of the gap are not so far from the CuO bond  
direction, if there are nodes at all in BSCCO. This is not in  
agreement with d-wave and more in favor of our model. Finally  
half-integer flux quantum effect has been observed in tricrystals  
of BSCCO \cite{kirtley}, providing rather convincing evidence  
for a change of sign of the order parameter on the Fermi surface.  
However this experiment does not provide a unique answer for  
the location of the nodes. For example, although the geometry  
is essentially the same as the one used for YBCO and is therefore  
consistent with d-wave, a g-wave order parameter 
$\Delta (\theta ) $  $\approx  $   
$\cos(4\theta ) $  is also a possible solution. More generally, since
the Josephson junctions  ( and in particular the scattering  
at the interfaces ) are not controlled in the tricrystals, it  
seems quite difficult to be secure about the position of the  
nodes from this kind of experiments. In conclusion it is fair  
to say that the problem of locating the nodes in BSCCO is not  
at all settled. At the same time it would clearly be a crucial  
piece of information to know in a reliable way this location  
since this would definitely allow to narrow the number of possible  
models and be another step toward identifying the mechanism  
of high $T_{c } $ superconductivity.  \par 

\section{HBCCO}
Let us consider now the case of the Hg compounds. Band structure  
calculations give a more complex situation than for BSCCO. Indeed  
the compound $HgBa_{2}CuO_{4+\delta } $ (Hg-1201) is found to have  
only a $CuO_{2} $ derived band crossing the Fermi 
level \cite{singh,rodr,novik,barbiel},   
although there is a HgO derived band which is coming quite near  
( typically 0.1 eV) to cross also the Fermi level from above.  
It is not clear that the precision of the calculations is enough  
to be certain about it. Moreover the role of oxygen doping,  
which is held responsible for the metallic and superconducting  
properties of this compound \cite{singh}, is also difficult  
to take into account and it seems to be quite important  \cite{singhpick}.  
Therefore it will be quite interesting to see what experiment  
will tell in this respect. The present tunneling data \cite{chen}  
are more in favor of a standard BCS density of states. This  
is coherent with our model since, with a single band, we expect  
only anisotropic s-wave pairing. However a two-band model is  
still an open possibility in this compound. An indication in  
this direction seems to be that, for Hg nuclei, $T_{1} $  has  
essentially the same value at  $T_{c } $ for Hg-1201 and Hg-1212,  
showing a similarity between these two compounds.  \par 
  \bigskip 
Indeed calculations in $HgBa_{2}CaCu_{2}O_{6+\delta } $ (Hg-1212)  
and $HgBa_{2}Ca_{2}Cu_{3}O_{8+\delta } $ (Hg-1223) 
find \cite{singhh,rodr,novik}   
the HgO derived band crossing also the Fermi level, in addition  
to the $CuO_{2} $  bands, which makes these compounds ''self-doped''.  
The situation is actually quite similar to the one found in  
BSCCO. The Fermi surface displays a small electron pocket centered  
at the X point. Because of hybridization between HgO and $CuO_{2} $  
 bands, the electronic states on this sheet of the Fermi surface  
have a mixed  HgO - $CuO_{2} $  character, as well as those of  
the nearby $CuO_{2} $ bands in the vicinity of the X point. An  
anticrossing is clearly seen. Therefore the necessary ingredients  
for our model are present, just as in BSCCO, and they can lead  
in the same way to nodes in the gap. Experimentally Raman scattering  
on Hg-1212 \cite{sacuto}gives evidence for nodes in the gap,  
but it shows also that some nodes (if not all) are away from  
the $45^{o}$ direction. This is in disagreement with d-wave and in  
agreement with our model. The low temperature dependence of  
the in-plane penetration depth in Hg-1223 \cite{panag}seems  
to be linear, which would indicate the existence of nodes in  
the gap. The available tunneling \cite{rossel} data are also  
suggestive of nodes. To summarize the present situation in these  
Hg compounds is coherent with our model.  \par 

\section{TBCCO}
We consider finally the case of the Tl compounds. Here again  
band structure calculations give results quite similar to the  
ones found for Hg compounds. There is a TlO band which crosses  
\cite{kasow,yu,haman,singht} the Fermi level 
for $Tl_{2}Ba_{2}Ca_{n-1}Cu_{n}O_{4+2n} $
 with n=1,2,3 and 4, giving rise to a small electron pocket  
for the Fermi surface. However for $Tl_{2}Ba_{2}CuO_{6} $  there  
is some disagreement between various calculations \cite{kasow,singht}  
and this result seems to be within the error bars. There are  
differences with Hg compounds. First the electron pocket is  
found around the $\Gamma  $  point. The states corresponding to this  
part of the Fermi surface are fairly delocalized throughout  
the cell with contributions not only from Tl and O atoms, but  
also from the apical O and even the $d(z^{2}) $ orbital of Cu  
atoms ( as a result, for $Tl_{2}Ba_{2}CuO_{6} $  \cite{singht},  
this piece of the Fermi surface has a more pronounced 3-dimensional  
character ). Then there is also the possible existence of hole  
pockets \cite{kasow,yu}, derived from flat $CuO_{2} $ bands. Finally  
since one has always the problem of dealing with oxygen doping,  
compounded with the experimental uncertainty on the atomic positions,  
the band structure situation is quite complicated.   \par 
  \bigskip 
We can again apply our model as in the case of BSCCO and say  
that Coulomb repulsion will lead to an order parameter with  
opposite signs on the sheets of the Fermi surface linked to  
the $CuO_{2} $  band and to the TlO band. This might explain the  
observation \cite{tsuei} of spontaneously generated half quantum  
flux in $Tl_{2}Ba_{2}CuO_{6+\delta } $  tricrystals. Nevertheless  
a major difference in the results of band structure calculations  
as compared with the Bi and Hg compounds is the lack of crossing  
of the $CuO_{2} $  and the TlO bands. Therefore in our model we  
would not expect nodes to appear. However it might very well  
be that the experimental situation is in this respect slightly  
different from the one predicted by band structure calculations  
\cite{kasow,singht} and that there is actually some crossing.  
This possibility has also been considered in the context of  
the inter-layer pairing mechanism \cite{schutz}. As we have  
mentionned there is enough uncertainty in the band structure  
calculations to allow for this possibility. Clearly direct information  
from ARPES, ACAR,dHvA experiments would be quite useful in order  
to settle this question. In our model these crossings would  
lead to nodes in the gap, in a way analogous to what we have  
found in BSCCO and HBCO. Actually available Raman scattering  
experiments \cite{nemet} favor the existence of nodes in a way  
very analogous to HBCO since they give, in $B_{1g} $ symmetry,  
a response linear in frequency at low $\omega , $ in the superconducting  
phase.  \par 

\section{NMR}
Since our model requires two kind of metallic bands in the Bi,  
Hg and Tl compounds, it appears important to find experimentally  
if we are indeed in this situation or not. This is actually  
rather difficult. We have already discussed this problem for  
BSCCO and found that there is no good experimental evidence  
against metallic BiO planes. Naturally one would rather like  
to find experiments proving definitely that they are indeed  
metallic, but we are not aware of any. Surprisingly the (quasi)  
quadratic structure of these compounds is in this respect a  
disadvantage, since in the case of YBCO the most reliable evidences  
of the metallic character of the chains come from the anisotropic  
properties linked to the orthorhombic structure. The difficulty  
comes from the microscopic nature of the information we are  
looking for, since for any macroscopic property we will not  
be able to know which one of the two bands is responsible for  
it.  \par 
  \bigskip 
Nevertheless it would seem that the NMR is an effective experiment  
in order to obtain this information since it is a local probe.  
The case of YBCO is a good example of this \cite{tachi}, although  
it seems to be somewhat forgotten. Indeed the superconducting  
transition is clearly seen as a strong drop both in the Knight  
shift K and in the relaxation rate $1/T_{1} $ of Cu(1) nuclei  
of the CuO chains, and the behaviour below $T_{c } $  is qualitatively  
similar to what is observed for Cu(2) in the $CuO_{2} $  planes.  
This shows that there is on this Cu(1) site a non vanishing  
amplitude for the wavefunction of the quasiparticles subject  
to the superconducting condensation. Moreover the quantitative  
temperature dependence for Cu(1) below $T_{c } $ is clearly different  
 from the one observed for the planes Cu(2), both for K and  
 $1/T_{1} $ . This shows that what is seen by Cu(1) nuclei is  
not merely a transferred part of the wavefunction of quasiparticles  
located dominantly in the $CuO_{2} $  planes, since in this case  
we would expect the temperature dependence to be the same for  
Cu(1) and Cu(2). Therefore the NMR favors a two-band picture  
for YBCO. This picture is also supported by the fact that the  
magnitude of $1/T_{1} $ at $T_{c } $  are quite similar for Cu(1)  
and Cu(2), and the same is true for the change in Knight shift  
between T = 0 and $T_{c } $  .  \par 
  \bigskip 
Let us consider the situation for Hg compounds. $^{199}Hg $  
NMR experiments have been performed recently 
\cite{suh}on $HgBa_{2}CuO_{4+\delta } $   
. Below $T_{c } $  they display a strong drop both in the Knight  
shift K and in the relaxation rate $1/T_{1} $ , so the Hg nuclei  
see the superconducting electrons. Moreover for $^{199}Hg $  
, $T_{1}T $ is essentially constant in the normal state ( the  
Knight shift is also constant with a ''Korringa ratio'' rather  
near what is expected from a Fermi liquid ), whereas $(T_{1}T)^{-1} $  
increases markedly with decreasing temperature for $^{63}Cu. $  
This is in favor of a two-band picture, just as in YBCO. One  
may argue that antiferromagnetic fluctuations are responsible  
for the increase in  $(^{63}T_{1}T)^{-1} $ and that these  
fluctuations completely cancel for symmetry reason at the Hg  
site, so that Hg has only a Korringa behaviour. However this  
implies an antiferromagnetic coupling between different $CuO_{2} $  
layers which is surprisingly strong ( taking in particular into  
account the disorder induced by oxygen doping in the Hg planes  
). A puzzling ingredient to this question is provided by the  
NMR on $HgBa_{2}CaCu_{2}O_{6+\delta } $ \cite{horv}where the temperature  
dependence of $(T_{1}T)^{-1} $ for $^{63}Cu $ and $^{199}Hg $  
is nearly the same. Clearly further experiments are needed to  
elucidate this point. Finally one has $^{199}T_{1} $  $\approx  $ $10^{2}. $  
 $^{63}T_{1} $  , but it is much more difficult to obtain firm  
conclusions from a comparison of the magnitude of $^{63}T_{1} $  
and $^{199}T_{1} $  since we deal with different nuclei, in  
contrast to YBCO.  \par 
  \bigskip 
We turn then to the Tl compounds. Again in $Tl_{2}Ba_{2}CuO_{6+\delta } $  
\cite{kitao,vyas}the Knight shift as well as the relaxation  
rate of $^{205}Tl $ display below $T_{c } $  a strong drop, which  
shows that the wavefunction of superconducting electrons extends  
in a quite sizeable way to Tl nuclei. One may wonder, as for  
the Hg compounds, if this is not due to a transferred hyperfine  
interaction of the Hg nuclei with electrons on the Cu site of  
the nearby $CuO_{2} $ planes through the apical oxygen. This possibility  
remains to be investigated in details. However, just as in the  
Mila and Rice phenomenological analysis \cite{mila} for YBCO,  
one expects this hyperfine coupling to occur through the contact  
interaction due to Tl-s orbitals and therefore to be isotropic.  
The other contributions should be much smaller. However the  
spin part of the Knight shift has a sizeable anisotropy ( $\approx  $  
30\% in a $T_{c } $  = 110K sample \cite{vyas} ), which is in contradiction  
with this hypothesis. Moreover, although it is difficult to  
obtain firm conclusions from magnitude   considerations, it  
is rather striking to see that the $^{63}Cu $ and $^{205}Tl $  
relaxation rates are nearly the same  ( $^{63}(1/T_{1}T) $ $\approx  $  
20 $sec^{-1}K^{-1} $  and  $^{205}(1/T_{1}T) $ $\approx  $ 8 $sec^{-1}K^{-1} $  
 in the normal state \cite{kitao,vyas}). Such a short $^{205}T_{1} $  
 is difficult to reconcile with a transferred hyperfine field.  
The natural interpretation of these $^{205}Tl $  NMR results  
is rather a confirmation of the two band picture given by band  
structure calculations.  \par 
  \bigskip 
Although more work is needed in order to reach a definite conclusion,  
we believe that the present NMR experimental evidence is leaning  
toward a two-band picture. However, even if we consider the  
single band hypothesis (assuming that only the $CuO_{2} $ band  
is relevant), the NMR shows unambiguously that the electronic  
wavefunction extends in a significant way to the Hg or Tl sites.  
Just as in our two band model, this provides a possible physical  
origin for an important Coulomb repulsion within this single  
band. This repulsion is likely to be quite anisotropic, since  
the probability of finding an electron on the Hg or Tl site  
will clearly change when one moves over the Fermi surface. This  
anisotropic repulsion might be responsible for a change of sign  
of the order parameter within this single band. Therefore in  
any case the NMR provides a clear hint that the Hg or Tl sites  
can not be safely omitted in a proper description of the electronic  
properties of the Hg or Tl compounds.  \par 

\section{NCCO  and  LSCO}
Let us finally consider the high Tc compounds, 
such as $La_{2-x }Sr_{x }CuO_{4} $   
or $Nd_{2-x }Ce_{x }CuO_{4} $ , which display only $CuO_{2} $  
planes without any other possible metallic component and hence  
provide a check for our physical picture. Indeed in this case  
we are naturally led to a single band and naturally band structure  
calculations lead in these compounds to a single $CuO_{2} $ derived  
band \cite{massne}. In the framework of our model we do not  
expect any nodes in the gap. This leads for these compounds  
to s-wave pairing, even if it is quite possible that the order  
parameter is fairly anisotropic. The experimental evidence in  
$Nd_{2-x }Ce_{x }CuO_{4} $  is presently clearly in favor of  
s-wave pairing. Indeed the low temperature dependence of the  
penetration depth follows the standard weak coupling behaviour  
\cite{wu}. Tunneling \cite{huang} gives also a fairly standard  
BCS shape, and moreover the Eliashberg function extracted from  
these tunneling data leads to a $T_{c } $ in good agreement with  
experiment, providing evidence for phonon mediated attractive  
interaction. Finally recent Raman scattering experiments \cite{stadlob}
go also in the direction of s-wave pairing since the results can  
be explained by an almost isotropic gap. Therefore NCCO seems  
in contradiction with the spin fluctuation mechanism. The fact  
that it is an ''electron doped'' compound, in contrast with all  
the other ''hole doped'' compounds is a rather unconvincing explanation  
for this discrepancy. Indeed one starts with a Fermi liquid  
with strong spin fluctuations and it is not clear at all why  
a small change in doping, allowing to go from the hole doped  
side to the electron doped side, would kill these spin fluctuations  
or reduce strongly their coupling to the electrons, while at  
the same time a sizeable coupling to phonons would appear.  \par 
  \bigskip 
On the other hand the experimental situation is 
unclear in $La_{2-x }Sr_{x }CuO_{4} $   
with some experiments in favor of s-wave and others
in favor of nodes. To our knowledge there is no phase sensitive  
experiments, nor precise measurements of the low temperature  
behaviour of the penetration depth. Recent Raman scattering  
experiments \cite{xkchen} have found results consistent with  
d-wave symmetry. However these results are not precise enough  
to prove convincingly the existence of nodes in the gap and  
some of their features do not agree with d-wave, but they display  
a clear anisotropy in the gap. Anyway, whatever the model, an  
experimental conclusion giving nodes in LSCO and no nodes in  
NCCO would be quite puzzling and we would rather hope for our  
understanding of high $T_{c } $ superconductivity that both LSCO  
and NCCO fall in the same class, whatever it may be.  \par 

\section{CONCLUSION}
In this paper we have proposed an extension to other high $T_{c } $  
 compounds of a model introduced earlier for YBCO. This extension  
relies on the fact that a number of these compounds are believed  
to be ''self-doped'', with a part of the compound giving holes  
to $CuO_{2} $  planes. Explicitely this doping part corresponds  
to the BiO, HgO, TlO planes in BSCCO, HBCCO, TBCCO respectively,  
just as it is the CuO chains in YBCO. This doping part is then  
considered as inert or unimportant in many models. We propose  
instead that it is metallic, which leads naturally to a multiband  
model. This picture is supported by band structure calculations.  
Naturally these calculations can not make sure that these doping  
bands are indeed metallic, but they make this possibility quite  
plausible. The final answer on this point can only be obtained  
from experiment. We can then take over the ingredients of our  
model in YBCO, and assume a repulsive pairing interaction between  
these doping bands and the $CuO_{2} $ bands. This leads to an  
opposite sign for the order parameter on these bands and to  
nodes whenever the Fermi surfaces of these bands cross. We have  
considered in particular the case of BSCCO and shown that this  
model accounts with reasonable parameters for the low temperature  
dependence of the penetration depth. An interesting feature  
of BSCCO is that the nodes are not located near the $45^{o}$ direction  
found in the d-wave model. This is in contrast with YBCO where  
the location is not far from $45^{o}$. Therefore the reliable experimental  
determination of the node locations is an important test for  
our model (we have discussed why ARPES does not provide a final  
answer in our view). We have then discussed the cases of HBCCO  
and TBCCO which have not been explored as much as BSCCO up to  
now. The situation in these compounds is somewhat similar to  
BSCCO with respect to our model. We have also considered the  
experimental evidence given by NMR and found that there are  
good indications for a metallic character of the doping bands  
in these last two compounds, although more experimental work  
is needed in order to get a definite anwer. Finally our model  
does not give nodes for NCCO and LSCO. This is in agreement  
with the present experimental situation for the first compound,  
but the evidence is quite ambiguous for the last one. We hope  
that experimental progress will allow fairly soon to confirm  
or invalidate our description. However whatever the answer we  
definitely need a consistent picture of all high $T_{c } $ superconductors,  
in agreement with all experimental data.  \par 

\section{ACKNOWLEDGEMENTS}
We are very grateful to H. Alloul, N. Bontemps, M. Cyrot, X.  
Leyronas, E. Maksimov, P. Mendels, P. Monod, and A. Trokiner for fruitful  
discussions.  \par 
\bigskip
* Laboratoire associ\'e au Centre National
de la Recherche Scientifique et aux Universit\'es Paris 6 et Paris 7.

\begin{figure}
\caption{Fermi surface in the Brillouin zone. The filled circles  
give the positions of the nodes of the gap.}
\label{Fig1}
\end{figure}

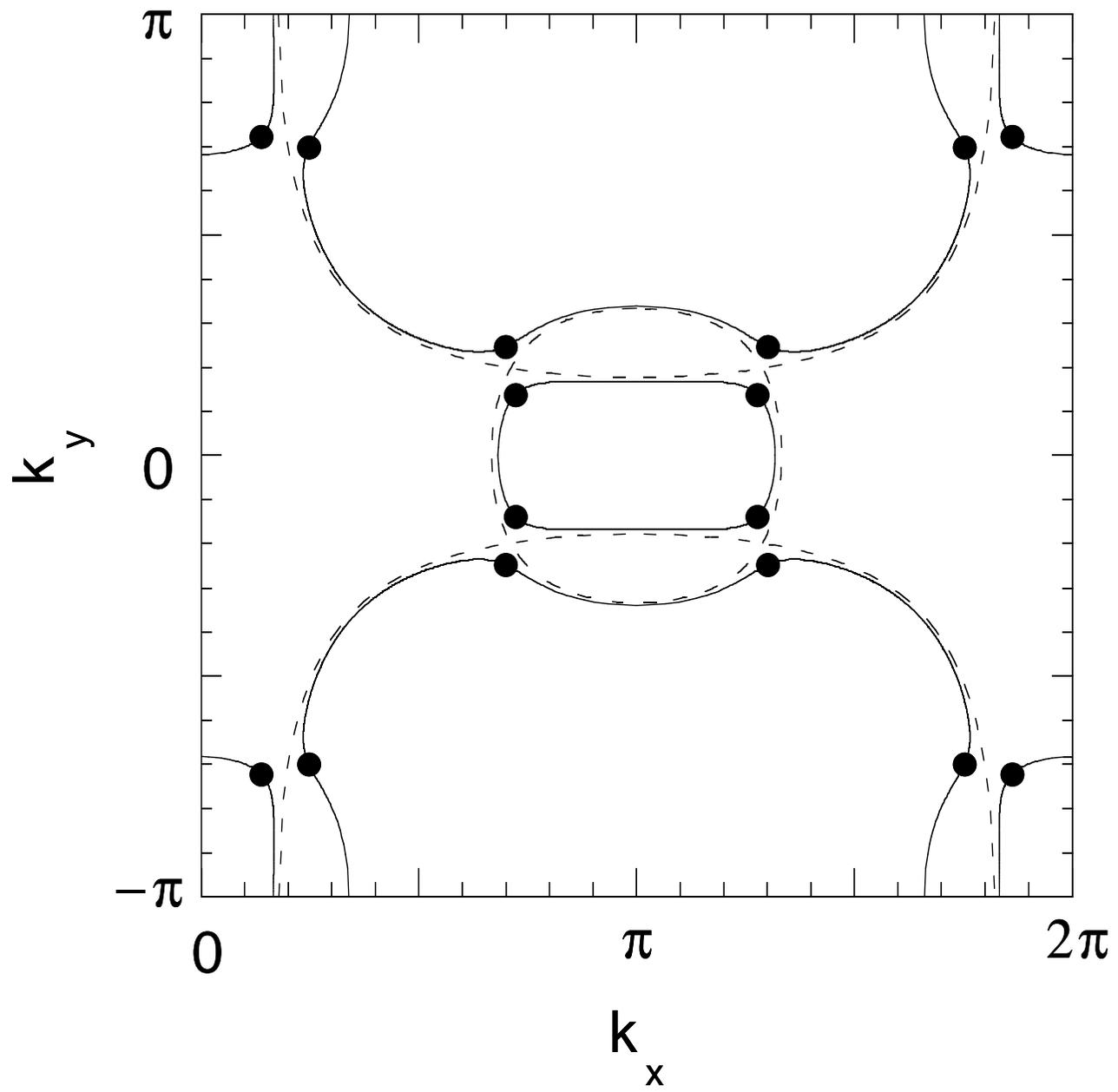
\end{document}